\documentclass{JAC2003}

\usepackage{enumitem}
\setlist{nolistsep}

\usepackage{graphicx}
\usepackage{booktabs}

\begin{document}
\title{COMPARISON OF THE CURRENT LHC COLLIMATORS AND THE SLAC PHASE 2  COLLIMATOR IMPEDANCES}

\author{H. Day\thanks{hugo.day@hep.manchester.ac.uk}, CERN, Switzerland, University of Manchester, UK and Cockcroft Institute, UK\\
F. Caspers, E. Metral, B. Salvant, CERN, Geneva, Switzerland\\
R.M. Jones, University of Manchester,  UK and Cockcroft Institute, UK}

\maketitle

\begin{abstract}
One of the key sources of transverse impedance in the LHC has been the secondary graphite collimators that sit close to the beam at all energies. This limits the stable bunch intensity due to transverse coupled-bunch instabilities and transverse mode coupling instability. To counteract this, new secondary collimators have been proposed for the phase II upgrade of the LHC collimation system. A number of designs based on different jaw materials and mechanical designs have been proposed. A comparison of the beam coupling impedance of these different designs derived from simulations are presented, with reference to the existing phase I secondary collimator design.
\end{abstract}

%
%
%
%
%

\section{INTRODUCTION}

As well as being a machine designed to push the forefront of high energy physics knowledge, the LHC has also presented a set of unique and demanding engineering and technical challenges to meet its design specifications. One of the most demanding of these requirements is a rigorous and comprehensive machine protection system. This is necessary due to the very large amount of stored energy within the LHC, both in terms of stored magnetic energy and beam energy (estimated at 10GJ and 350MJ at nominal intensity \cite{mach-protect}), which if not properly controlled can easily cause significant damage to the accelerator's components.

Due to the stringent mechanical requirements on the physical component of the collimation system it was decided to focus of physical robustness for the choice of jaw material. This led to the choice of carbon-reinforced carbon as the jaw material due to it's mechanical stability. However, this material is not ideal from the perspective of beam cleaning or beam coupling impedance \cite{imp-coll, imp-phase-2}. To achieve optimal performance for beam performance a number of phase 2 collimator designs have been proposed to address these concerns.


\section{CERN PHASE 2 CANDIDATE DESIGNS}

For the phase 2 upgrade of the LHC collimation system 3 priorities were chosen to guide the designs studied. These were (in not particular order); 

\begin{enumerate}
\item{Improved cleaning efficiency through the use of a high-Z material for the jaw material}
\item{Faster and more accurate collimator setup}
\item{Reduced beam coupling impedance}
\end{enumerate}

\begin{figure}
\begin{center}
\includegraphics[width=0.8\linewidth]{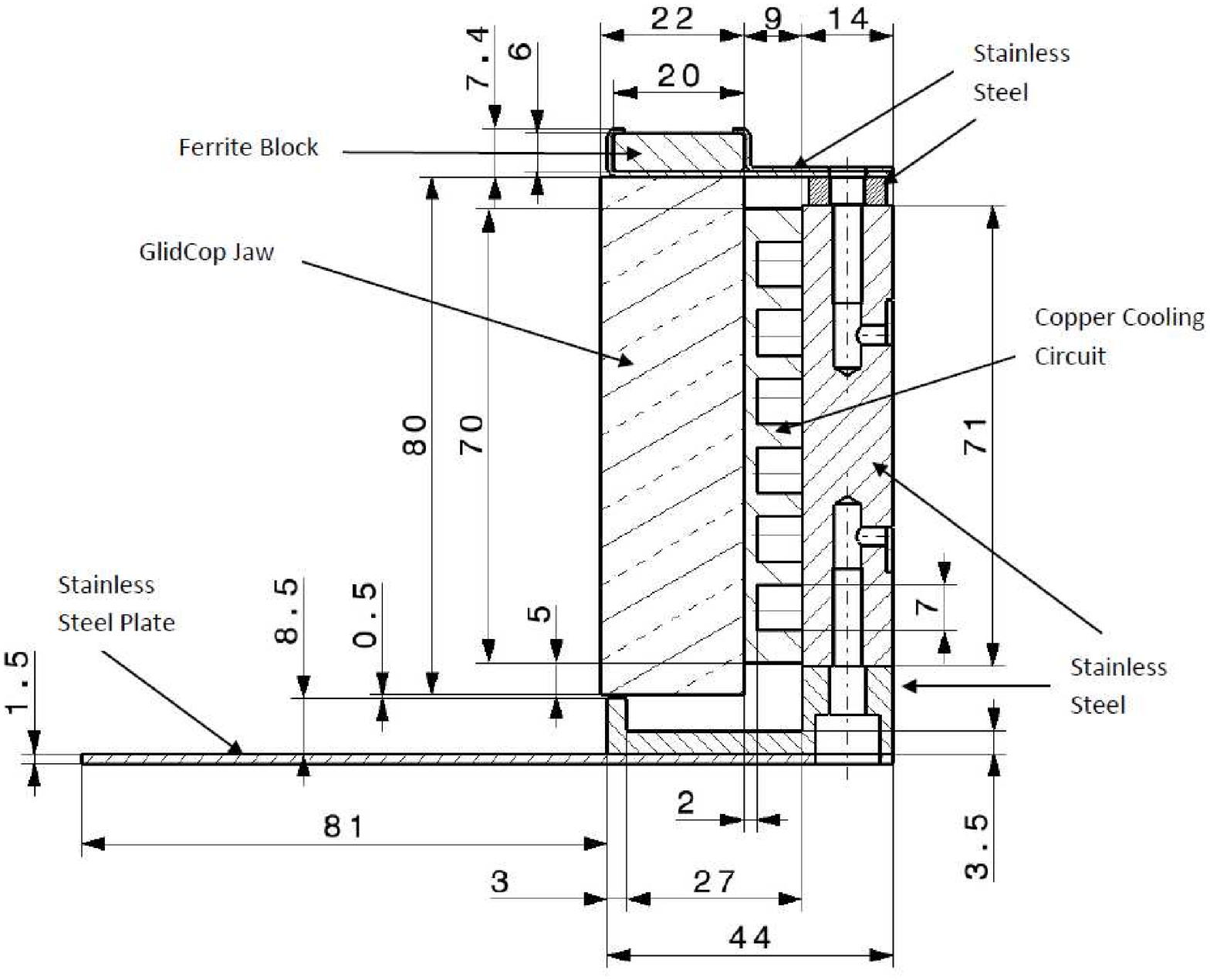}
\end{center}
\caption{CERN phase 2 design. The glidcop jaw material maybe altered depending on the optimal material for the design.}
\label{fig:cern-design}
\end{figure}

The second requirement is intended to be satisfied by the use of button BPMs integrated in the collimators themselves, whilst the first and third requirements require a careful choice of the jaw material and the geometry of the collimator. For the CERN design it was decided to have a jaw design which could have a variety of jaw materials applied (see Fig.~\ref{fig:cern-design}). A variety of possible jaw materials had been suggested, including metals, metal/diamond and dielectric materials the details of some of which are given in Table~\ref{tab:mat}. The materials can be grouped into metals, poor conductors (graphite) and dielectrics (Silicon Carbide (SiC))

\begin{table}[hbt]
\begin{center}
\caption{Material Conductive Properties}
\begin{tabular}{ | l | r | }
\hline
Material & Conductivity $S m^{-1}$ \\ \hline
Graphite & $7 \times 10^{4}$  \\ \hline
Glidcop & $5.33 \times 10^{7}$  \\ \hline
Molybdenum (Mo)& $1.76 \times 10^{7}$  \\ \hline
Mo Carbon-Diamond &  $4.5 \times 10^{6}$  \\ \hline
\end{tabular}
\end{center}
\label{tab:mat}
\end{table}


From the impedance point of view the main questions are what the optimal jaw material is and whether it is possible to screen a low-conductivity material with a high-conductivity material. The second of these points is due to mechanical considerations which may make a material that is not the optimal impedance choice a favoured material in terms of robustness and cleaning efficiency. To this end five material combinations were examined; 

\begin{enumerate}
\item{Glidcop jaw}
\item{Molybdenum jaw}
\item{Molybdenum carbon-diamond composite coated in 2mm of molybdenum}
\item{SiC tiles mounted in copper}
\item{graphite to give a comparison to the phase 1 collimators}
\end{enumerate}

\begin{figure}
\begin{center}
\includegraphics[width=0.8\linewidth]{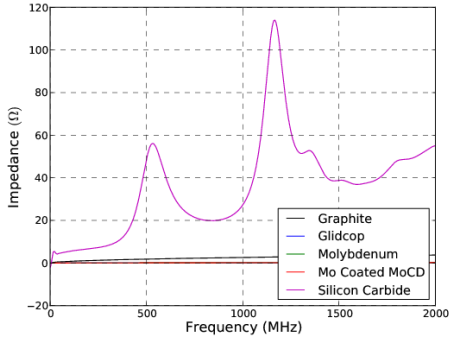}
\end{center}
\caption{Real part of the longitudinal impedance obtained by simulations off all candidate materials for the CERN phase 2 design. A jaw spacing of 6mm and a cross-section length of 0.2m was used}
\label{fig:cern-imp-all}
\end{figure}

To simulate the structure in Fig.~\ref{fig:cern-design}, CST Particle Studio \cite{cst-ref} was used. Two jaws of seperation 6mm and a length of 0.2m was used. An integrated wakelength of 20m was used. The mesh count for this model was $\sim$1.2 million cells, increasing slightly for models with highly lossy materials. As seen in Fig.~\ref{fig:cern-imp-all}, SiC produces a significantly larger real part of the longitudinal impedance that the existing graphite design and the metal/metal-composite designs. This is due to the large imaginary component of its complex permittivity. Similar results are observed for the imaginary longitudinal and transverse impedances also. As such, it can be determined that exposed SiC is not a suitable jaw materials.

\begin{figure}
\begin{center}
\includegraphics[width=0.8\linewidth]{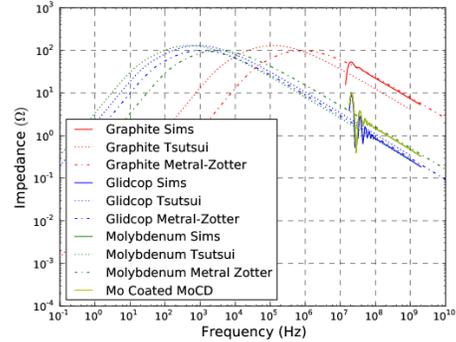}
\end{center}
\caption{Real part of the vertical dipolar impedance from simulations of a number of CERN designs compared to different resistive wall impedance models. Here the impedance is simulated for a beam displaced by 0.5mm from the centre axis with two jaws seperated by 6mm and 0.2m in length}
\label{fig:cern-trans}
\end{figure}

To allow for a comparison of transverse impedances of different jaw materials at low frequencies (in the range of kHz), it was decided to investigate whether it would be possible to use an existing analytical model to accurately represent the impedance of the design. This is due to the large wakelengths that would be necessary to carry out simulations that would cover low frequencies. Comparisons were made to the Tsutsui-model \cite{tsutsui-dip} and the Metral-Zotter model \cite{mz-trans} of resistive wall transverse dipole impedance. As can be seen in Fig.~\ref{fig:cern-trans}, there is good agreement between the Metral-Zotter model and the simulations for all materials over a large frequency range, from 20MHz up to 2GHz, with the results diverging at low frequencies due to an insufficiently long simulation of the wakelength. This indicates that the impedance in the CERN design is dominated by resistive wall impedances. Also notable is that the pure molybdenum jaw and the molybdenum coated molybdenum carbon-diamond jaws exhibit the same impedance profile demonstrating that the lower conductivity metal carbon-diamond composites are effectively screened by a coating of pure metal.

\section{SLAC PHASE 2 CANDIDATE}

\begin{figure}
\begin{center}
\includegraphics[width=0.8\linewidth]{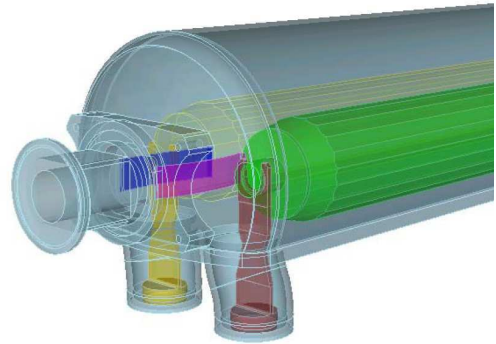}
\end{center}
\caption{SLAC Phase 2 collimator design. Image used courtesy of the SLAC design team.}
\label{fig:slac-geo}
\end{figure}

The SLAC design proposal takes a novel approach to ensuring long lasting jaw lifetime by using a rotating collimator concept \cite{slac-rot}. These jaws, made of copper, will have 20 sides with one acting as the collimator jaw at any one time. The complete geometry can be seen in Fig.~\ref{fig:slac-geo}. 

The design of the SLAC collimator gives two possible sources of impedance, both resistive wall impedances due to the jaw material and to cavity impedance from the surrounding vacuum tank. Previous work on the cavity modes within the vacuum tank has considered possible heating of the collimators \cite{slac-long}, but not the beam coupling impedance. A model with two different jaw spacings (2mm and 60mm) has been imported into CST Particle Studio \cite{cst-ref} and the beam coupling impedance simulated. These are compared to the impedance of the CERN design in the next section.

\section{COMPARISON BETWEEN THE DESIGNS}

\begin{figure}
\begin{center}
\includegraphics[width=0.8\linewidth]{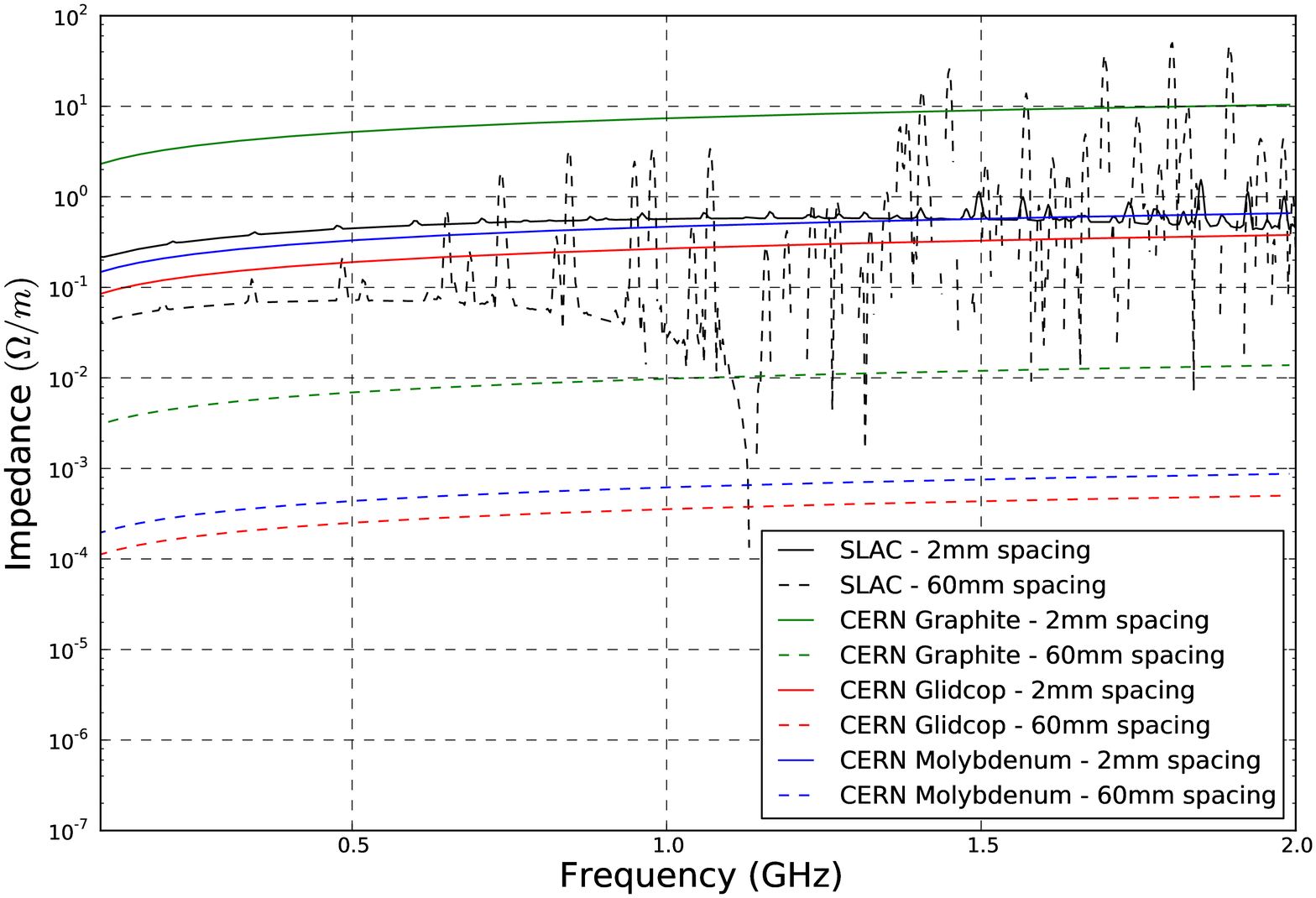}
\end{center}
\caption{Real part of the longitudinal impedance per unit length of the SLAC and CERN phase 2 designs at jaw spacing of 2mm and 60mm}
\label{fig:cern-slac-long-comp}
\end{figure}

The longitudinal impedance of the two designs is comparable across much of the frequency range. However, significant peaks are seen at frequencies above $\sim$500MHz for the SLAC design with 60mm jaw spacing (see Fig.~\ref{fig:cern-slac-long-comp}). This is due to the large number of large Q-factor trapped modes that occur above this frequency \cite{slac-long}. With a jaw spacing of 2mm these modes occur at much higher frequencies. However it can be seen that all designs give a better impedance profile than the equivalent graphite collimator across much of the frequency range considered. 
\begin{figure}
\begin{center}
\includegraphics[width=0.8\linewidth]{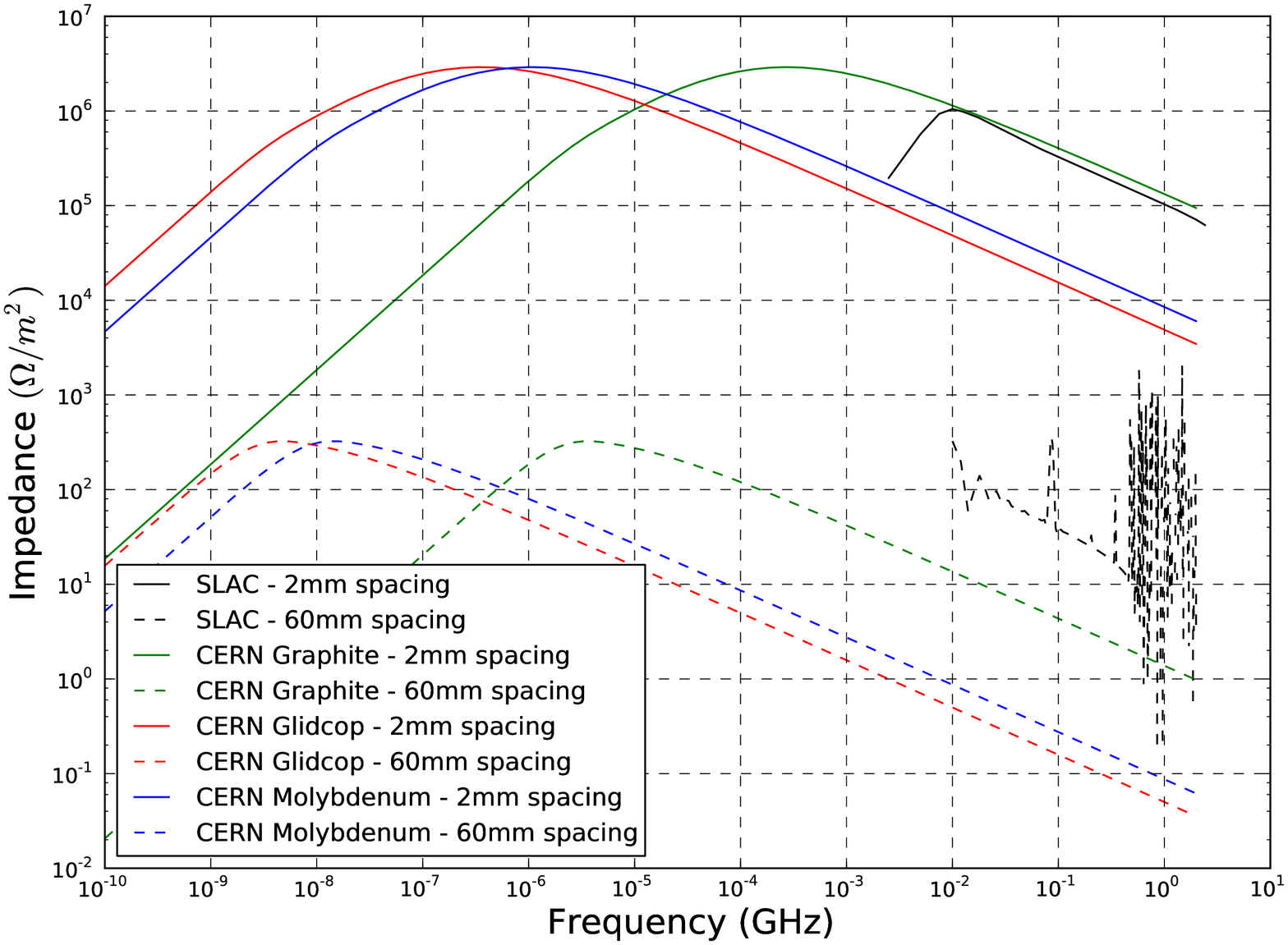}
\end{center}
\caption{Real part of the horizontal dipolar impedance per unit length of the SLAC and CERN phase 2 designs at jaw spacing of 2mm and 60mm}
\label{fig:cern-slac-trans-comp}
\end{figure}

Considering the real horizontal dipolar impedance (see Fig.~\ref{fig:cern-slac-trans-comp}) it can be seen that for a small jaw spacing the SLAC design demonstrates a much larger impedance than the CERN design. This is due to the SLAC design having a significantly larger cavity space and thus a large R/Q, whilst the CERN design is dominated by resistive wall impedances at this setting, and copper and glidcop having very similar conducitivities. Both the molybdenum and glidcop/copper based designs give significantly better impedance performance when compared to the graphite collimators for frequencies above $\sim$10kHz at a 2mm jaw spacing. This is significant due to the first unstable betatron line failling on 8kHz \cite{imp-coll}. For wider jaw spacings the performance only becomes more favourable to the metallic designs.

\section{CONCLUSION}

The impedance simulations demonstrate that the use of metals and metal coated metal-carbon/diamond composites in the CERN phase 2 design produces better impedance conditions than that of the existing graphite collimators. Conversely, unexposed silicon carbide produces significantly worse impedance conditions due to it's high permittivity across all frequencies. Compared to the CERN design, the SLAC design provides comparable impedance profiles for small jaw spacings. However, for large jaw spacings it has a number of large impedance peaks at frequencies above 1GHz. This would require further investigation to clarify their effect on beam operation.

\section{ACKNOWLEDGEMENTS}
Thanks to Liling Xiao, Jeff Smith, Tom Markeiwicz and Steve Lundgren for providing details of the previous studies of the SLAC phase 2 design . Thanks to Nicola Mariani, Alessandro Bertarelli and Alessandro Dallacchio for providing the materials data for the CERN design jaw materials and schematics.

\end{document}